\documentclass[proc]{edpsmath}
\usepackage{graphicx}

\begin{document}

\title{Exponential wealth distribution: \\
a new approach from functional iteration theory}
\thanks{R.L-R. acknowledges financial support from  the Spanish research project
 DGICYT-FIS2009-13364-C02-01.} 
\author{Ricardo L\'opez-Ruiz}\address{Dep. of Computer Science \& BIFI, Faculty of Science, 
Universidad de Zaragoza, Zaragoza, Spain; \email{rilopez@unizar.es}}
\author{Jos\'e-Luis L\'opez}\address{Dep. of Mathematical and Informatics Engineering, 
Universidad P\'ublica de Navarra, Pamplona, Spain; \email{jl.lopez@unavarra.es}}
\author{Xavier Calbet}\address{BIFI, Inst. de Biocomput. y F\'{\i}sica de Sistemas Complejos,
Universidad de Zaragoza, Zaragoza, Spain; \email{xcalbet@googlemail.com}}
\dedicated{\it The Authors dedicate this paper, in memoriam, \\
to Jos\'e F\'elix S\'aenz Lorenzo, the former Director of the BIFI Institute.} 
\begin{abstract}
Exponential distribution is ubiquitous in the framework of multi-agent systems. 
Usually, it appears as an equilibrium state in the asymptotic time evolution of
statistical systems. It has been explained from very different perspectives.
In statistical physics, it is obtained from the principle of maximum entropy \cite{jaynes1957}.
In the same context, it can also be derived without any consideration about information theory,
only from geometrical arguments under  the hypothesis of equiprobability in phase space \cite{lopez2008}. 
Also, several multi-agent economic models based on mappings, with random, deterministic or 
chaotic interactions, can give rise to the asymptotic appearance of the exponential wealth distribution 
\cite{estevez2008,pellicer2010}. An alternative approach to this problem
in the framework of iterations in the space of distributions has been presented in \cite{lopez2010}.
Concretely, the new iteration given by
\begin{equation}
f_{n+1}(x) = \int\!\!\int_{u+v>x}\,{f_n(u)f_n(v)\over u+v} \;  {\mathrm d}u{\mathrm d}v \,. \nonumber
\label{syst1}
\end{equation}
\noindent It is found that the exponential distribution is a stable fixed point of
the former functional iteration equation. From this point of view, it is easily 
understood why the exponential wealth distribution (or by extension, other kind of distributions) 
is asymptotically obtained in different multi-agent economic models \cite{yako2009}. 
\end{abstract}
\begin{resume} 
Diff\'erentes approches pour d\'eriver le r\'egime asymptotique exponentiel  dans les syst\`emes 
statistiques sont possibles. Ici une nouvelle \'equation fonctionnelle inspir\'ee 
dans les syst\`emes \'economiques se propose pour expliquer la distribution exponentielle. 
Dans ce cas, cette distribution est le seul point fixe auquel la dynamique de cette \'equation 
fonctionnelle \'evolue quand l'it\'eration va vers l'infini. 
De ce point de vue, c'est facile de comprendre l'ubiquit\'e 
de cette distribution (ou d'autres) en diff\'erents probl\`emes statistiques r\'eels. 
\end{resume}
\maketitle

\newpage
\section*{Introduction}

If someone were to make the similarity of a gas of particles with human society we certainly 
would think that it is a ridiculous option. If this comparison is restricted only to economic aspects,
maybe the simile is not so disproportionate. Let us think, for instance, that human or economic agents
exchange money or wealth in commercial transactions in the same way that particles in a gas exchange 
energy in collisions. It is clear that we could always argue 
that economic agents are endowed with intelligence and a certain foresight and determination 
in their decisions to purchase and sale, characteristics not present in the random interactions of 
particles in a gas. But even ignoring these details about the intelligence of the agents, 
the similarity between markets and gases has been successfully introduced in recent years. 
It has managed to reproduce surprisingly some of the features observed in the real economy, 
such as for example the distribution of wealth in society. This approach that applies 
methods of statistical physics to describe the economy is what has been called 
Econophysics \cite{chakra2010}. Likewise, models that compare the behavior of the economy 
in a system of trading agents with which occurs with energy or 
other quantities in a gas of particles are called gas-type models \cite{yakoven2009}.

\begin{figure}[h]
  \centering \includegraphics[width=0.6\hsize]{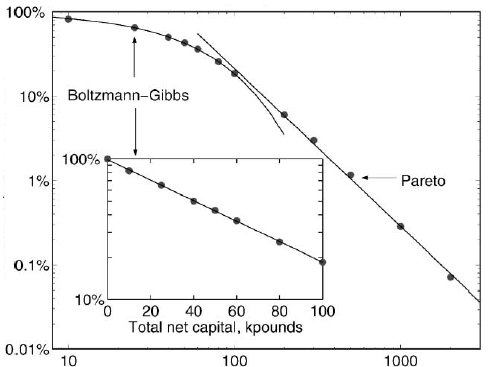} 
  \caption{Cumulative probability distribution of the total net capital (wealth) in kpounds 
  in UK (data for 1996) shown in log-log, and log-linear the inset. 
  The exponential (Boltzmann-Gibbs) and power (Pareto) 
  laws fit for the 95\% and 5\% of the population, respectively. 
  (Figure extracted from Ref. \cite{dragu2001}). }
  \label{Fig1}
\end{figure}

Nowadays it is known that the society of the Western (capitalist) economies can be divided into two 
distinct groups if we follow the distribution of wealth (wages, property, etc.) \cite{dragu2001}. 
On the one hand, the 95\% of the population, the middle and lower economic classes of society, 
allocate their wealth according to an exponential distribution. Furthermore, there is a group 
of 5\% of individuals, those with privileged incomes, whose wealth is distributed according to 
a power law distribution. (See Figure \ref{Fig1}).

\begin{figure}[h]
  \centering \includegraphics[width=0.6\hsize]{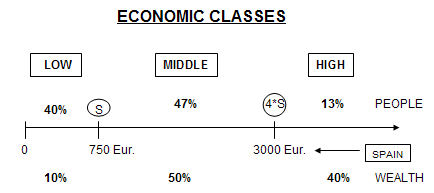} 
  \caption{Economic partition generated in the society by an exponential wealth distribution.
  The minimum salary $S$ is considered as the reference level for the low economic class,
  and $4S$ is taken as the reference level for the high economic class. The average salary
  in this case will be $2S$. (Figure presented in \cite{lopez2010}). }
  \label{Fig2}
\end{figure}

In this work, we are interested in an alternative explanation of the exponential 
behavior of the wealth distribution in western societies. 
First, let us see what an exponential distribution of income means.
If we assume for example that in Spain the average salary per month is $1500$ euros, 
we can consider the low incomes those salaries below the half of the average, i.e. 
$750$ euros per month, and the high incomes those revenues above the double 
of the average, i.e. $3000$ euros per month. 
Then, the middle economic classes are formed by those individuals with incomes 
in the range between $750$ and $3000$ euros per month. If now we know that wealth 
is distributed in an exponential way , it means, roughly speaking,  
that the $40\%$ of the population are low economic class and receive 
the $10\%$ of the total payroll, the $47\%$ of the population are middle economic class 
and pocket the $50\%$ of the total salary mass , and finally the remaining $13\%$ are high 
economic class and allocate the $40\%$ of the total wealth. (See Figure \ref{Fig2}).

\section{Estrategies explaining the exponential behavior in statistical systems}

Different strategies can generate the exponential wealth distribution with a gas-type model. 
One of them is as follows \cite{dragu2000}: (1) a set of agents is placed each with the same initial 
amount of money, (2) transactions between agents are allowed in a random manner, 
thus every time two agents are randomly chosen, (3) they put their money all together and they share
it in a random way too, (5) the gas evolves with this type of interactions and the 
asymptotic wealth distribution appears to be the exponential distribution. Note that in 
this model all the components are random, the pairs of interacting agents at each moment 
and the way they divide their money. As local interactions conserve the money, 
the global dynamics is also conservative and the total amount of money is constant in time. 

\begin{figure}[h]
  \centering \includegraphics[width=0.8\hsize]{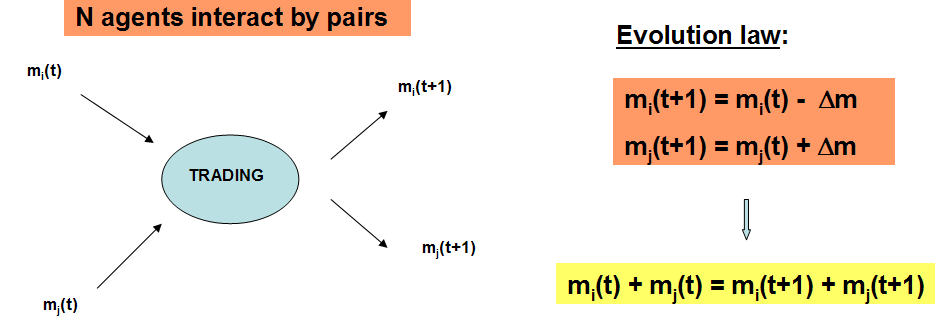} 
  \caption{Visual representation of the trading mechanism between pairs $(m_i,m_j)$
  of economic agents in the gas-like models. The time $t$ is discrete.}
  \label{Fig3}
\end{figure}

Then the time evolution of this system in phase space is made on the hyperplane 
defined by the constraint that imposes the conservation of the total amount of money. 
If we suppose that none of the points of the hyperplane is privileged over any other one,
i.e. if the equiprobability hypothesis is assumed, then it can be newly calculated that 
the distribution of wealth in such a system is exponential. 
Let us remark that this method (presented in \cite{lopez2008}) derives the exponential 
distribution by simple geometric arguments. This strategy is in some way related with the 
variational method proposed by Jaynes in Ref. \cite{jaynes1957}, which obtains the exponential
distribution in a statistical system as the result of maximizing the entropy of the system 
when the average value of the energy (or other variables) in the ensemble is fixed. 
Although equiprobability seems to be strongly related to the exponential behavior of the measurable
variables of a system, it is not mandatory that equiprobability be reached by random mechanisms.
In fact, the exponential distribution has also been found in multi-agent economic systems
with just strictly deterministic elements \cite{estevez2008} or even in gas-like models
with chaotic ingredients \cite{pellicer2010} .

\section{Another explanation for the exponential distribution: \\ a functional iteration model}

If we assume an ensemble of $N$ economic agents trading with each other as explained in
the former section, the evolution equations of this gas-like model can be written as
\begin{eqnarray}
m'_i &=& \epsilon \; (m_i + m_j), \nonumber\\
m'_j &=& (1 - \epsilon) (m_i + m_j), \label{model1}\\
i , j &=& 1 \ldots N, \nonumber
\end{eqnarray}
where $\epsilon$ is a random number in the interval $(0,1)$. 
The agents $(i,j)$ are randomly chosen and they trade with money (or goods, commodities, etc.)
in such a way that their initial money, $(m_i, m_j)$, at time $t$ is transformed 
in $(m'_i, m'_j)$ at time $(t+1)$. It is found that the asymptotic distribution
$p_f$, obtained by numerical simulations, is the exponential (Boltzmann-Gibbs) 
distribution \cite{dragu2000},
\begin{equation}
p_f(m)=\beta \exp(-\beta \,m), \hspace{0.5cm}\hbox{with}\hspace{0.5cm}\beta={1/ \langle m \rangle},
\label{eq-exp}
\end{equation}
where $p_f(m) {\mathrm d}m$ denotes the PDF (probability density function), i.e.
the probability of finding an agent with money (or energy in a gas system) between 
$m$ and $m + {\mathrm d}m$. 
Evidently, this PDF is normalized, $\parallel p_f\parallel=\int_0^{\infty} p_f(m){\mathrm d}m=1$. 
The mean value of the wealth, $\langle m \rangle$, can be easily calculated by
$\langle m \rangle=\sum_i m_i/N$.   

\begin{figure}[h]
  \centering \includegraphics[width=0.8\hsize]{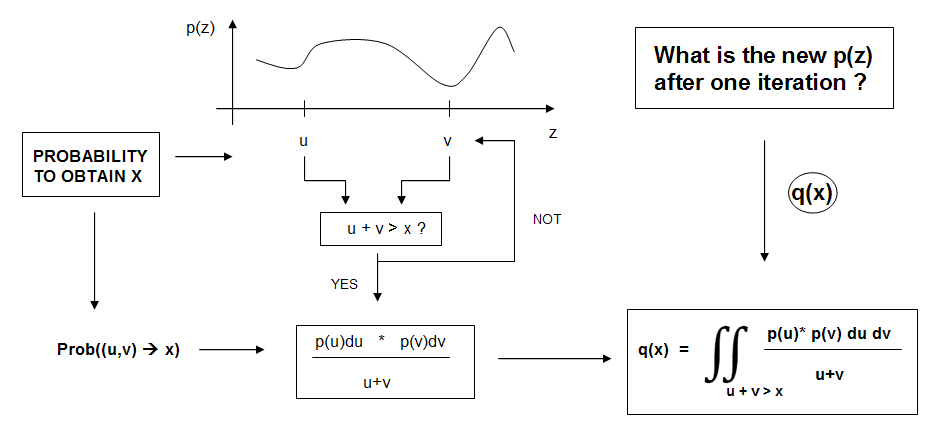} 
  \caption{Derivation of the nonlinear evolution operator $\cal T$ for the 
  agent-based economic model given by Eqs. (\ref{model1}). See explanation in the text.}
  \label{Fig4}
\end{figure}

We can regard this model as an evolution problem in the space of PDFs, such as
\begin{equation}
\lim_{n\rightarrow\infty} {\cal T}^n \left(p_0(m)\right) \rightarrow p_f(m),
\label{eq-operT}
\end{equation}
where an initial wealth distribution $p_0(m)$ with a mean wealth value $\langle p_0 \rangle$
evolves in time under the action of an operator $\cal T$ to asymptotically reach 
the equilibrium distribution $p_f(m)$, that in this particular case is the exponential one
which presents the same average value $\langle p_f \rangle=\langle p_0 \rangle$. 
This is a macroscopic interpretation of Eqs. (\ref{model1}) in the sense that,
under this point of view, each iteration of the operator $\cal T$ means 
that many interactions, of order $N/2$,
have taken place between different pair of agents. If the subindex $n$
indicates the time evolution of $\cal T$, we can roughly assume that $t\approx N*n/2$,
where $t$ follows the microscopic evolution of the individual tradings (or collisions) 
between the agents (this alternative microscopic interpretation can be seen in \cite{calbet2011}).

Thus the question to answer is if it is possible to find $\cal T$ and, if this is
the case, how $\cal T$ looks like. Now, we proceed to derive  $\cal T$ in an easy way
(see Fig. \ref{Fig4}, presented in \cite{lopez2010}). Suppose that $p_n$ is the wealth
distribution in the ensemble at time $n$. 
The probability that two agents with money $(u,v)$ interact will be
$p_n(u)*p_n(v)$. As the trading is totally random, their exchange can give rise with equal
probability to any value $x$ comprised in the interval $(0,u+v)$. Then, the probability to
obtain a particular $x$ (with $x<u+v$) for the interacting pair $(u,v)$ will be 
$p_n(u)*p_n(v)/(u+v)$. Finally we can obtain the probability to have money $x$ at time $(n+1)$.
It will be the sum of the probabilities for all the pairs of agents $(u,v)$ able to generate the
quantity $x$, that is, all the pairs verifying $u+v>x$. $\cal T$ has then the form
of a nonlinear integral operator,
\begin{equation}
p_{n+1}(x)={\cal T}p_n(x) = \int\!\!\int_{u+v>x}\,{p_n(u)p_n(v)\over u+v} 
\; {\mathrm d}u{\mathrm d}v \,.
\label{eq-T}
\end{equation}

\begin{figure}[h]
  \centering \includegraphics[width=0.8\hsize]{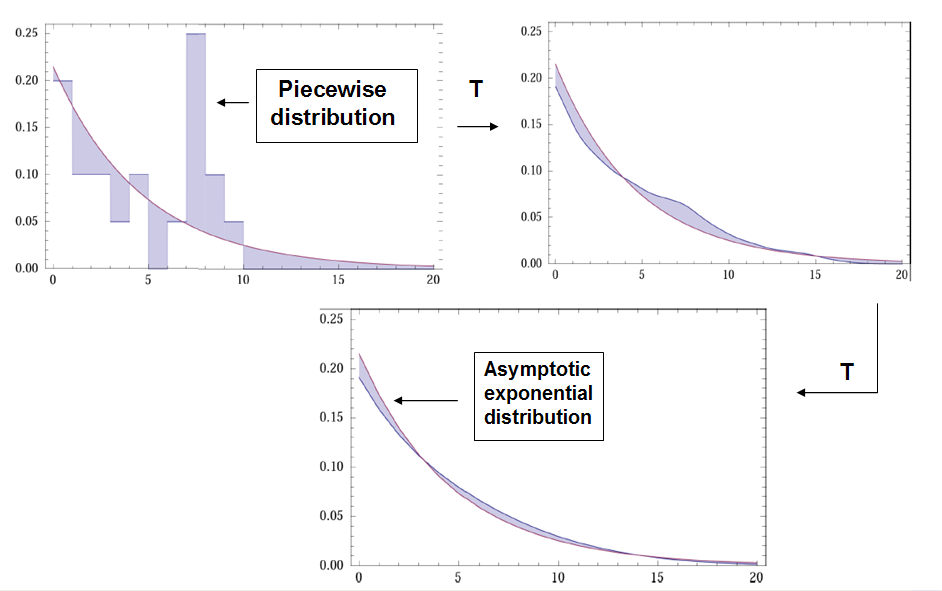} 
  \caption{An example of the action of $\cal T$ in the space of PDFs. In this case, $p_0(x)$ 
  is chosen as a normalized piecewise distribution. Observe that after two iterations of $\cal T$
  the system has practically reached the asymptotic state, i.e. the exponential distribution $p_f(x)$.
  (Figure avanced from \cite{shiva2011}). }
  \label{Fig5}
\end{figure}

If we suppose $\cal T$ acting in the PDFs space, it can be proved that $\cal T$ conserves 
the norm ($\parallel ·\parallel$), i.e. $\cal T$ maintains the total richness of the system, 
$\parallel{\cal T} p\parallel=\parallel p\parallel=1$. 
By extension, it also conserves the mean wealth of the system, 
$\langle {\cal T}p \rangle=\langle p \rangle$. 
Then, it can be seen that the exponential distribution $p_f(x)$ with the right average value
is the steady state of $\cal T$, i.e. ${\cal T}p_f=p_f$.
Also it can be proved that this exponential distribution $p_f$ is the only fixed point 
of $\cal T$. In consequence, the relation (\ref{eq-operT}) is true. Finally, it is also found that 
the entropy is always an increasing quantity with time. All these properties will be developed 
and shown in next papers \cite{lopez2011,shiva2011}.

\section{Conclusions}

From an economic point of view, different approaches to obtain the exponential (Boltzmann-Gibbs)
distribution have been recalled. The gas-like models that interpret the market as an ensemble of
agents trading with money in the same way as particles exchange energy in a gas can help to
derive the asymptotic Gibbs regime. Following this insight, a new operator
in the framework of functional iteration theory has been proposed. The uniparametric family 
of exponential distributions is the only family of attractive fixed points of this operator. 
Then, this model explains in a straightforward way the ubiquity of the exponential distribution 
in the asymptotic regime of the more diverse problems, those with some economic inspiration included.


\end{document}